\def\tg{\hbox{tg}}

\def\ln{\ell{n}}
\documentstyle[12pt]{article}
\begin{document}
\begin{titlepage} \vspace{0.2in} \begin{flushright}
MITH-97/06 \\ \end{flushright} \vspace*{1.5cm}
\begin{center} {\LARGE \bf  Neutrino masses and mixings\\} 
\vspace*{0.8cm}
{\bf She-Sheng Xue$^{\star}$\\}
INFN, Section of Milan, Via Celoria 16, Milan, Italy\\
Physics Department, University of Milan, Italy\\
\vspace*{1.5cm}
{\bf   Abstract  \\ } \end{center} \indent
 
We propose a novel theoretical understanding of neutrino masses and mixings,
which is attributed to the intrinsic vector-like feature of the regularized
Standard Model at short distances. We try to explain the smallness of Dirac
neutrino masses and the decoupling of the right-handed neutrino as a free
particle. Neutrino masses and mixing angles are completely related to each
other in the Schwinger-Dyson equations for their self-energy functions. The
solutions to these equations and a possible pattern of masses and mixings are
discussed. 

\vfill \begin{flushleft} May, 1997   \vspace*{1.5cm} \\
\noindent{\rule[-.3cm]{5cm}{.02cm}} \\
\vspace*{0.2cm} \hspace*{0.5cm} ${}^{\star}$ 
E-mail address: xue@milano.infn.it\end{flushleft} \end{titlepage}

\noindent
{\bf 1.}\hskip0.3cm
Since their appearance, neutrinos have always been extremely peculiar. Their
charge neutrality, near masslessness, flavour mixing, and parity-violating
coupling have been at the centre of a conceptual elaboration\cite{yu,moh} and
an intensive experimental analysis\cite{je} that have played a major role in
donating to mankind the beauty of the Standard Model (SM). In the
present letter, we propose a novel theoretical understanding of neutrino
masses and mixings in a left-right symmetric extension of the SM, inspired by
the intrinsic vector-like feature of chiral gauge theories at high-energies. 

The notion of the vector-like feature stands for: given any conserved quantum
numbers of chiral gauge symmetries of a regularized quantum field theory, there
must be the exactly equal numbers of left-handed and right-handed fermions,
and parity-conserving gauge couplings at short distances\footnote{This feature
is generic, not only for the lattice regularization, since the ``no-go''
theorem\cite{nn} that proves this feature actually is based on the argument of
chiral gauge anomalies.}. The feature is clearly phenomenologically
unacceptable. However, it really implies right-handed neutrinos and
parity-conserving gauge theories in short distances\footnote{Although the
solution to this ``no-go'' theorem has not been completely found.}. This calls for
the left-right symmetric model ($SU_L(2)\otimes SU_R(2)\otimes
U_{B-L}(1)$)\cite{lr}, where parity is unbroken at high energies and its
nonconservation at low-energies occurs through a spontaneous symmetry breakdown
mechanism. 

The left-right symmetric extension that we suggest still possesses
$SU_L(2)\otimes U_Y(1)$ gauge symmetries. The right-handed
doublets are assigned to bound three-fermion states ($i=e,\mu,\tau$):
\begin{equation}
\left(\matrix{\nu^3_i\cr i^3}\right)_R;\hskip0.2cm
\nu^3_{iR}\sim (\bar \nu_R\cdot \nu^i_L)\nu_R,\hskip0.2cm i^3_R\sim 
(\bar i_R\cdot i_L)i_R,
\label{threeb}
\end{equation}
where $\nu_R$ is a gauge singlet and a unique right-handed neutrino for three
families. These three-fermion states (\ref{threeb}) carry the appropriate
quantum numbers of the $SU_L(2)\otimes U_Y(1)$ gauge symmetries so that the 
SM gauge symmetries are preserved. We do not need new elementary gauge and fermion
fields for the $SU_R(2)$ sector. The neutrino fields (both elementary and
composite (\ref{threeb})) considered in this letter are purely Weyl neutrinos. 

These three-fermion states can be formed\cite{xue97} by effective high-dimension operators
of fermionic fields, $L_{h.d.o.}$, which 
are due to the underlying physics at the cutoff $\Lambda$, 
\begin{equation}
L_{\rm effective}=L_{SM}+L_{h.d.o.}.
\label{eff}
\end{equation}
Before knowing what the dynamics of underlying physics is, {\it a priori}, we
conceive that the possibilities for $L_{h.d.o.}$ are those allowed by the gauge
symmetries of the SM or other unification models,
e.g., $SO(10)$\cite{so10,ss}. 

In order to consistently achieve parity-violation at low-energies, we 
postulate that at an intermediate energy-threshold $\epsilon$: 
\begin{equation}
250GeV \ll\epsilon < \Lambda,
\label{epsilon}
\end{equation}
the three-fermion states (\ref{threeb}) dissolve into their constituents,
i.e.~turn into the virtual states of their constituents (three-fermion cuts).
This is due to the vanishing of the binding energy of three-fermion states
(\ref{threeb}) at the threshold $\epsilon$ (\ref{epsilon})\cite{xue97}. This 
phenomenon
could be realized on the basis that the effective high-dimension operators
for binding the three-fermion states are
momentum-dependent. However, it is difficult to demonstrate 
this dynamics, since it relates to a non-perturbative issue of finding the 
inferred fix point and determining the spectrum and
relevant high-dimension operators that realize the symmetries. It
should be pointed out that this phenomenon is not spontaneous symmetry
breaking and no Goldstone bosons occur. 

In this letter, we consider the above extension of the SM as a model instead of
a demonstrated theory. For the purpose of studying neutrino masses and mixings, we only discuss the
three-fermion state (\ref{threeb}) and disregard the $U_{B-L}(1)$
symmetry. Presumably, the spectrum of right-handed three-fermion states should
be much richer than (\ref{threeb}), so as to preserve chiral gauge symmetries
of a unified theory. 

\vskip0.3cm
\noindent
{\bf 2.}\hskip0.3cm
$\nu_R$ couples to other fermions via $L_{h.d.o.}$ in (\ref{eff}). The 1PI
(one-particle irreducible) vertices between $\nu_R$ and $\nu_L^i$ give Dirac
neutrino masses. Rather than the Sea-Saw mechanism\cite{ss}, we discuss another
possible dynamics for decoupling $\nu_R$ and small Dirac neutrino masses. 

We assume, the couplings of $\nu_R$ low-frequency
modes ($p\ll\Lambda$) to other fermions are suppressed by
$\left({p\over\Lambda}\right)^2$. This is to mean that every external
$\nu_R$ line (with external momentum $p$) of any 1PI-vertices is associated
with $\left({p\over\Lambda}\right)^2$. Thus, at low-energies, $\nu_R$
decouples from other fermions, as a result, Dirac neutrino masses are very
small and $\nu_R$ is a free particle. On the other hand, $\nu_R$ high-frequency
modes (250GeV$\ll\epsilon<p<\Lambda$) couple strongly enough to other 
fermions to form three-fermion states (\ref{threeb}). 

This is equivalent to assuming that $L_{h.d.o.}$ (\ref{eff}) possess
the following $\nu_R$ shift-symmetry\cite{gp},
\begin{equation}
\nu_R(x) \rightarrow \nu_R(x)+\delta,
\label{shift}
\end{equation}
where $\delta$ is a constant. The decoupling of $\nu_R$ can be
shown by the Ward identities of this shift-symmetry\cite{xue97},
\begin{equation}
\gamma_\mu\partial^\mu\nu'_R(x)
+\left({\partial_\mu\over\Lambda}\right)^2\langle \hat O(x)\rangle
-{\delta\Gamma\over\delta\bar\nu'_R(x)}=0,
\label{w}
\end{equation}
where ``$\Gamma$'' is the effective potential with non-vanishing external
sources ($J,\eta$); the external field $\nu_R'\equiv\langle\nu_R\rangle$, and
$\langle \hat O(x)\rangle$ is the expectation value of the operator $\hat O$
w.r.t.~the generating functional $Z[J,\eta]$. Based on this Ward identity, we can
get all 1PI vertices containing at least one external $\nu'_R$. 

The first are the self-energy functions $\Sigma_{\nu_i}(p)$ ($i=e,\mu,
\tau$) of neutrinos. Performing a functional derivative of eq.~(\ref{w})
w.r.t.~the primed field $\nu'^i_L(0)$ and then putting external sources 
$\eta=J=0$, we obtain,
\begin{equation}
\left({\partial_\mu\over\Lambda}\right)^2\langle \hat O^i(x)\rangle_\circ-
{\delta^2\Gamma\over\delta\nu'^i_L(0)
\delta\bar\nu'_R(x)}=0,
\label{ws1}
\end{equation}
where $\langle\hat O^i(x)\rangle_\circ\equiv\langle\hat O^i(x)\rangle|_{J=\eta=0}
$, and as a result,
\begin{equation}
\int_xe^{-ipx}{\delta^2\Gamma\over\delta\nu'^i_L(0)
\delta\bar\nu'_R(x)}={1\over2}\Sigma^i(p)= 
\left({p^2_\mu\over\Lambda^2}\right)\langle\hat O^i(0)\rangle_\circ.
\label{small}
\end{equation}
For low energies $p\ll\Lambda$, Dirac neutrino masses $\Sigma_{\nu_i}(p)\ll 1$. 

The second is the wave renormalization function. The 
functional derivative of eq.(\ref{w}) w.r.t.~$\nu'_R(0)$
and then $\eta=J=0$ lead to,
\begin{equation}
(\gamma_\mu P_R)^{\beta\alpha}\partial^\mu \delta(x)
-{\delta^2\Gamma\over\delta\nu'^\alpha_R(0)
\delta\bar\nu'^\beta_R(x)}=0.
\label{wf}
\end{equation}
The two-point function is then given by,
\begin{equation}
\int_xe^{-ipx}
{\delta^{(2)}\Gamma\over\delta\nu'_R(x)\delta\bar\nu'_R(0)}=i
\gamma_\mu p^\mu,\label{free}
\end{equation}
indicating that $\nu_R$ does not receive wave-function renormalization $Z_3$.

The third are the $n$-point ($n>2$) 1PI interacting vertices. 
Analogously, we can obtain,
\begin{equation}
{\delta^{(n)}\Gamma\over\delta^{(n-1)}(\cdot\cdot\cdot)\delta\bar\nu'_R(x)}
\sim O\left({p^2_\mu\over\Lambda^2}\right),\hskip0.3cm n>2.\label{n}
\end{equation}
where $\delta^{(n-1)}$ indicates $(n-1)$ derivatives 
w.r.t.~other external fields. 

These three identities, eqs.(\ref{small},\ref{free}) and (\ref{n}),
show us two conclusions owing to the $\nu_R$ shift-symmetry (or
only $\nu_R$ high-frequence modes coupling to other fermions):
(i) the Dirac neutrino masses due to high-dimension operators ($L_{h.d.o.}$)
are extremely small;
(ii)
the right-handed neutrino $\nu_R$ at low-energies is a free particle
and decouples from other physical particles. 

\vskip0.3cm
\noindent
{\bf 3.}\hskip0.3cm
Once the soft spontaneous symmetry breaking occurs at the weak scale,
the right-handed fermion states are the mixed states comprising
the elementary state $\nu_R$ ($i_R$) and the composite state $\nu^3_{iR}$ 
($i^3_R$) ($i=e,\mu,\tau$):
\begin{equation}
\Psi^{\nu_i}_R=(\nu_R,\nu^3_{iR});\hskip0.5cm \Psi^i_R=(i_R,i^3_R).
\label{mixing}
\end{equation}
The self-energy functions $\Sigma_{\nu_i}(p)$ ($\Sigma_i(p)$) of neutrinos
(charged leptons) are couplings between $\nu^i_L(i_L)$ and mixing right-handed
fermion states $\Psi^{\nu_i}_R(\Psi^i_R)$, rather than the couplings between 
only elementary states $\nu^i_L(i_L)$ and $\nu_R(i_R)$ in the SM. This will
become very clear in the Schwinger-Dyson (SD) equations for the 
self-energy functions. 

For the reason that the three-fermion states (\ref{threeb}) carry the $SU_L(2)$
quantum number, there must be an interacting 1PI vertex between $W^\pm,Z^\circ$
bosons
and composite right-handed fermions (\ref{threeb}) in the high-energy region.
We may write this effective 1PI coupling for $W^\pm$ as, 
\begin{eqnarray}
\Gamma^{ij}_\mu (q)&=&i{g_2\over2\sqrt{2}}V_{ij}\gamma_\mu (P_L+f(q))\label{wv}\\
f(q)&\not=&0,\hskip0.5cm q\ge\epsilon,
\end{eqnarray}
where $g_2$ is the $SU_L(2)$ coupling and $V_{ij}$ is the CKM matrix\cite{data},
and that for the $Z^\circ$ is similar.
At the energy threshold (\ref{epsilon}), where the 
three-fermion states dissolve into their constituents,
the effective 1PI vertex function $f(q)$ must vanish,
\begin{equation}
f(q)|_{q\rightarrow\epsilon+0^+}\rightarrow 0,
\label{threshold}
\end{equation}
which results in the parity-violating gauge-couplings in low-energies.

In $L_{h.o.d.}$ of the effective lagrangian (\ref{eff}), we may have 
following gauge-invariant operators interacting between up quarks 
($q^u_i=u,c,t$) and neutrinos, and between down quarks ($q^d_i=d,s,b$) and 
charged leptons ($i=e,\mu,\tau$),
\begin{eqnarray}
&&G\bar\psi^\beta_{iL}(x)\cdot 
\left[{\partial^2_\mu\over\Lambda^2}\nu_R(x)\right]\bar q^u_{iR}(x)
\cdot Q^\beta_{iL}(x),
\label{tau1}\\
&&G\bar\psi^\beta_{iL}(x)\cdot i_R(x)\bar q^d_{iR}(x)\cdot Q^\beta_{iL}(x),
\label{tau1'}
\end{eqnarray}
where $\psi^\beta_{iL}$ and $Q^\beta_{iL}$ ($\beta=1,2$) are the $SU_L(2)$
doublet of leptons and quarks respectively. Once quarks are massive, these 
operators are the sources
providing explicit chiral symmetry breaking to generate lepton masses. 

Turning off all gauge interactions and putting
the four-fermion coupling\footnote{The same strength $G$ of all four-fermion
couplings is assigned for a potential unification.} $G$ to its critical
value $G\rightarrow 4+0^+$ given by the $\bar tt$-condensate model\cite{tt}, 
we obtain the simplest SD gap-equations at the cutoff ($p=\Lambda$),
\begin{eqnarray}
\Sigma^\circ_{\nu_i}(p)&=&\left({p^2\over\Lambda^2}\right)\Sigma_{q^u_i}
(\Lambda),\label{su5'}\\
\Sigma^\circ_i(\Lambda) &=& \Sigma_{q^d_i}(\Lambda),
\label{su5}
\end{eqnarray}
where we are henceforth in the basis of mass eigenstates. 
Eq.(\ref{su5'}) shows that the neutrino $\Sigma^\circ_{\nu_i}(p)$ decouples from
the quark $\Sigma_{q^u_i}(\Lambda)$ for $p\ll\Lambda$, which agrees with 
(\ref{small}). Eq.(\ref{su5}) is reminiscent of the predictions in the 
$SU(5)$ unification model\cite{su5}. 

It should be pointed out that there are other possible gauge-invariant
four-fermion interactions and their corresponding tadpole diagrams contribute
to eqs.(\ref{su5'},\ref{su5}) as well. We consider all such contributions
to eqs.(\ref{su5'},\ref{su5}) as lepton's {\it bare masses} at the cutoff,
which are actually explicit symmetry breaking terms in the full SD equations (\ref{gapn},
\ref{gapl}). 

\vskip0.3cm
\noindent
{\bf 4.}\hskip0.3cm
Turning on all gauge interactions, we study the full SD equations for the 
lepton self-energy functions. In the rainbow approximation and the Landau gauge, 
these equations can be written as,
\begin{eqnarray}
\Sigma_{\nu_i}(p)&=&\Sigma^\circ_{\nu_i}(p)+Z_{\nu_i}(p)+W_{\nu_i}(p),
\label{gapn}\\
\Sigma_i(p)&=&\Sigma^\circ_i(\Lambda)+Z^1_i(p)+Z^3_i(p)+W_i(p)+\gamma_i(p),
\label{gapl}
\end{eqnarray}
where
\begin{eqnarray}
\gamma_i(p)&=&3e^2\int^\Lambda_{p'} {1\over (p-p')^2}
{\Sigma_{i}(p')\over p'^2+\Sigma_{i}^2(p'^2)}\label{g}\\
Z^1_i(p)&=&3\lambda^2\int^\Lambda_{p'} {1\over (p-p')^2+M^2_z}
{\Sigma_{i}(p')\over p'^2+\Sigma_{i}^2(p'^2)},\label{z1}
\end{eqnarray}
$\lambda=g_2\tg\theta_w(\sin^2\theta_w-{1\over2})$ and $\theta_w$ is
the Weinberg angle.

Because of the effective 1PI coupling (\ref{wv}) and three-fermion states
(\ref{threeb}) above the threshold $\epsilon$ (\ref{epsilon}), $W^\pm$ and
$Z^\circ$ bosons contribute to eqs.(\ref{gapn},\ref{gapl}). These most peculiar
contributions are $W_{\nu_i}(p), Z_{\nu_i}(p)$
and $W_i(p),Z^3_i(p)$: 
\begin{eqnarray}
W_{\nu_i}(p)&=&\left({g_2\over2\sqrt{2}}\right)^2
|V_{ij}|^2\int_{|p'|\ge\epsilon}^\Lambda {f(p'-p)\over (p-p')^2+M_w^2}
{\Sigma_{j}(p'^2)\over p'^2+\Sigma_{j}^2(p'^2)},\label{tw}\\
Z_{\nu_i}(p)&=&\left({g_2\over2\cos\theta_w}\right)^2
\int_{|p'|\ge\epsilon}^\Lambda {f(p'-p)\over (p-p')^2+M_z^2}
{\Sigma_{\nu_i}(p'^2)\over p'^2+\Sigma_{\nu_i}^2(p'^2)};\label{tz}\\
W_i(p)&=&\left({g_2\over2\sqrt{2}}\right)^2
|V_{ij}|^2\int_{|p'|\ge\epsilon}^\Lambda {f(p'-p)\over (p-p')^2+M_w^2}
{\Sigma_{\nu_j}(p'^2)\over p'^2+\Sigma_{\nu_j}^2(p'^2)},\label{bw}\\
Z^3_i(p)&=&\left({g_2\cos2\theta_w\over2\cos\theta_w}\right)^2
\int_{|p'|\ge\epsilon}^\Lambda {f(p'-p)\over (p-p')^2+M_z^2}
{\Sigma_i(p'^2)\over p'^2+\Sigma_i^2(p'^2)},\label{bz}
\end{eqnarray}
where the integration of the internal momentum $p'$
starts from the intermediate threshold $\epsilon$ to the cut-off $\Lambda$.

We note that eqs.(\ref{su5'},\ref{su5},\ref{g},\ref{z1}) are the 1PI couplings
between elementary left-handed and right-handed fields, whereas 
eqs.(\ref{tw})-(\ref{bz}) are the 1PI couplings
between elementary left-handed fields and right-handed three-fermion fields 
(\ref{threeb}). Thus, we clarify that the full self-energy functions 
$\Sigma_{\nu_i}(p)$ (\ref{gapn}) and $\Sigma_i(p)$ (\ref{gapl}) are
the 1PI couplings between elementary left-handed fields and mixed right-handed
fields (\ref{mixing}). At low-energies, external momenta $p\ll\epsilon$, 
$\Sigma_{\nu_i}(p)$ (\ref{gapn}) are very small, because eqs.(\ref{su5'}),
(\ref{tw}) and (\ref{tz}) turn to zero, these latter are due to disappearance 
of three-fermion states and the 1PI-vertex (\ref{threshold}).

\vskip0.3cm
\noindent
{\bf 5.}\hskip0.3cm
We are in the position to solve the SD eqs.(\ref{gapn},\ref{gapl}). Assuming the 
scale $\epsilon$ is large enough and $p'>\epsilon\gg 1$, 
we approximate the inhomogeneous terms (\ref{tw},\ref{bw}) to be,
\begin{equation}
W_{\nu_i}(p)\simeq\alpha_w(p)|V_{ij}|^2\Sigma_j(\Lambda),\hskip0.2cm
W_i(p)\simeq\alpha_w(p)|V_{ij}|^2\Sigma_{\nu_j}(\Lambda),
\label{www}
\end{equation}
where
\begin{equation}
\alpha_w(p)\simeq\left({g_2\over2\sqrt{2}}\right)^2
\int^\Lambda_{|p'|\ge\epsilon} {f(p'-p)\over (p-p')^2+M_w^2}
{1\over p'^2}.
\label{alpha}
\end{equation}

In the high-energy region ($x=p^2>\epsilon\gg 1$), where $M^2_w,M^2_z$ and 
nonlinearity are negligible, the SD integral equations (\ref{gapn},\ref{gapl}) 
can be converted to the following boundary value problems\footnote{Analogous 
technique can be found in ref.\cite{kogut}}, 
\begin{eqnarray}
{d\over dx}\left(x^2\Sigma'_{\nu_i}(x)\right)+{f_n\over 4}
\Sigma_{\nu_i}(x)&=&0,
\label{ndeqb'}\\
\Lambda^2\Sigma'_{\nu_i}(\Lambda^2)+\Sigma_{\nu_i}(\Lambda^2)&=&
\Sigma^\circ_{\nu_i}(\Lambda)+\alpha_w(\Lambda)|V_{ij}|^2\Sigma_j(\Lambda);
\label{nboundaryb'}\\
{d\over dx}\left(x^2\Sigma'_i(x)\right)+{f_c\over 4}
\Sigma_i(x)&=&0,
\label{deqb'}\\
\Lambda^2\Sigma'_i(\Lambda^2)+\Sigma_i(\Lambda^2)&=&
\Sigma^\circ_i(\Lambda)+\alpha_w(\Lambda)|V_{ij}|^2\Sigma_{\nu_j}(\Lambda),
\label{boundaryb'}
\end{eqnarray}
where $f_n,f_c$ are perturbative functions of electroweak couplings.
These are differential equations with the inhomogeneous
boundary conditions at the cutoff. 
The generic solutions to eqs.(\ref{ndeqb'},\ref{deqb'}) for $(x\gg 1)$ 
are given by,
\begin{eqnarray}
\Sigma_{\nu_i}(x) &\simeq& {A_{\nu_i}\mu^2\over\sqrt{ x}}{\rm sinh}
\left({1\over2}\sqrt{
1-f_n}\ln({x\over\mu^2})\right),
\label{nsolution}\\
\Sigma_i(x) &\simeq& {A_i\mu^2\over\sqrt{ x}}{\rm sinh}
\left({1\over2}\sqrt{
1-f_c}\ln({x\over\mu^2})\right),
\label{lsolution}
\end{eqnarray}
where $A_{\nu_i}, A_i$ are arbitrary constants and $\mu$ is an inferred scale.
Substituting (\ref{nsolution}) into (\ref{nboundaryb'}) and 
(\ref{lsolution}) into (\ref{boundaryb'}) in the low-energy limit 
($\mu\ll\Lambda$), we obtain the gap-equations:
\begin{eqnarray}
\alpha_w(\Lambda)|V_{ij}|^2\Sigma_j(\Lambda)&=&{1\over2}\Sigma_{\nu_i}(\Lambda)
+{1\over2}\sqrt{1-f_n}\Sigma_{\nu_i}(\Lambda)
-\Sigma^\circ_{\nu_i}(\Lambda),\label{nlow3b'}\\
\alpha_w(\Lambda)|V_{ij}|^2\Sigma_{\nu_j}(\Lambda)&=&{1\over2}\Sigma_i(\Lambda)
+{1\over2}\sqrt{1-f_c}\Sigma_i(\Lambda)
-\Sigma^\circ_i(\Lambda).
\label{low3b'}
\end{eqnarray}
Since the lepton bare masses (\ref{su5'},\ref{su5}) are defined 
when all gauge interactions are turned off, we can rewrite the RHS of 
the gap-equations (\ref{nlow3b'},\ref{low3b'}),
\begin{eqnarray}
\alpha_w(\Lambda)|V_{ij}|^2\Sigma_j(\Lambda)&=&-{f_n\over4}
\Sigma_{\nu_i}(\Lambda),
\label{neutri}\\
\alpha_w(\Lambda)|V_{ij}|^2\Sigma_{\nu_j}(\Lambda)&=&
-{f_c\over4}\Sigma_i(\Lambda).
\label{low3b''}
\end{eqnarray}

As a consequence of the gap-equations (\ref{nlow3b'},\ref{low3b'}), the 
lepton self-energy functions must be non-trivial 
\begin{equation}
\Sigma_{\nu_i}(\Lambda)\not=0;\hskip0.3cm {\rm and } \hskip0.3cm 
\Sigma_j(\Lambda)\not=0,
\end{equation}
if quarks are massive. The gap-equations
(\ref{su5}) strongly imply that the hierarchical pattern of charged lepton
masses is mainly due to the hierarchical pattern of down quark masses. 
Eqs.(\ref{neutri},\ref{low3b''}) show that the pattern of neutrino masses is
determined by the CKM-mixing angles and charged lepton masses. 

\vskip0.3cm
\noindent
{\bf 6.}\hskip0.3cm
The six gap-equations (\ref{neutri},\ref{low3b''})
relate neutrino and charged lepton masses at the cutoff. Noticing the fact
that 
$\Sigma(p)$ must be continuous functions of ``$p$'' from $p=\Lambda$ to 
$p\ll\epsilon$ for the locality of quantum field theories, and 
the ratios of $\Sigma(p)$'s in the same charge sector 
(but different generations) should
be scale invariant (renormalization group invariant), we take ratios between
the two equations of eqs.(\ref{neutri}), the two equations of eqs.(\ref{low3b''})
respectively and scale them down to the low-energy scale. We end up with
four independent equations: 
\begin{eqnarray}
{m_{\nu_e}\over m_{\nu_\mu}}&=&{|V_{\nu_ee}|^2m_e-|V_{\nu_e\mu}|^2m_\mu
+|V_{\nu_e\tau}|^2m_\tau\over
|V_{{\nu_\mu}e}|^2m_e-|V_{{\nu_\mu}\mu}|^2m_\mu
+|V_{{\nu_\mu}\tau}|^2m_\tau},\label{r1}\\
{m_{\nu_\mu}\over m_{\nu_\tau}}&=&{|V_{{\nu_\mu}e}|^2m_e
-|V_{{\nu_\mu}\mu}|^2m_\mu
+|V_{{\nu_\mu}\tau}|^2m_\tau\over
|V_{{\nu_\tau}e}|^2m_e-|V_{{\nu_\tau}\mu}|^2m_\mu
+|V_{{\nu_\tau}\tau}|^2m_\tau},\label{r1'}
\end{eqnarray}
and
\begin{eqnarray}
{m_e\over m_\mu}&=&{|V_{e\nu_e}|^2m_{\nu_e}-|V_{e{\nu_\mu}}|^2m_{\nu_\mu}
+|V_{e{\nu_\tau}}|^2m_{\nu_\tau}\over
|V_{\mu\nu_e}|^2m_{\nu_e}-|V_{\mu{\nu_\mu}}|^2m_{\nu_\mu}
+|V_{\mu{\nu_\tau}}|^2m_{\nu_\tau}},\label{r3}\\
{m_\mu\over m_\tau}&=&{|V_{\mu\nu_e}|^2m_{\nu_e}-|V_{\mu{\nu_\mu}}|^2m_{\nu_\mu}
+|V_{\mu{\nu_\tau}}|^2m_{\nu_\tau}\over
|V_{\tau\nu_e}|^2m_{\nu_e}-|V_{\tau{\nu_\mu}}|^2m_{\nu_\mu}
+|V_{\tau{\nu_\tau}}|^2m_{\nu_\tau}}.\label{r3'}
\end{eqnarray}
In these equations, all fermion masses are defined at the same low-energy
scale. We make an appropriate chiral rotation in the second family and all
fermion masses are positive. Given charged lepton masses, four relationships
satisfied by four mixing angles and three neutrino masses, which are no longer
seven free parameters. These equations give a class of solutions for
the possible patterns of neutrino masses and mixing angles. 

Setting $\theta_{13}=0$ so that $\theta_{13}$ and $\delta_{13}$ decouple from
eqs.(\ref{r1}-\ref{r3'}), and assuming ${m_{\nu_e}\ll m_{\nu_\mu}\ll m_{\nu_\tau}}$, 
we get a possible pattern:
\begin{eqnarray}
\tg^2\theta_{12}&=&{m_e\over m_\mu}+O\left({m_{\nu_e}\over m_{\nu_\tau}}\right),
\hskip0.2cm m_{\nu_e}\sim \sin^2\theta_{13}=0,
\label{12}\\
\sin^2\theta_{23}&=&{m_\mu+m_e\over
m_e+m_\mu+m_\tau}+ O\left({m_{\nu_\mu}\over m_{\nu_\tau}}\right),
\hskip0.2cm {m_{\nu_\mu}\over m_{\nu_\tau}}\simeq 5.8\cdot 10^{-3}.
\label{23}
\end{eqnarray}
This coincides with the ``standard'' scenario of neutrino masses and mixings
that can also be realized by the sea-saw mechanism in GUT models\cite{yu,moh}.
Given this hierarchical pattern and $m_{\nu_\tau}\sim O$(eV) for the desired
HDM of the Universe, the results (\ref{12})-(\ref{23}) are consistent with the
small angle MSW solution to the solar neutrino problem, and however we may need
an addition sterile neutrino $\nu_s$\cite{yu} to explain the atmospheric
neutrino deficit. 

Many other possible patterns of neutrino masses and mixing are discussed in
the literature\cite{yu}, phenomenologically based on cosmological constraints and
three neutrino experiments (the solar neutrinos, atmospheric neutrino, LSND).
While, on the other hand, these neutrino experiments need to be
substantiated\cite{je}. 

At this point, we should emphasize that in this extension of the SM, the four
CKM mixing angles ($\theta_{ij}$) are totally extrinsic elements, qualifying
the true pattern of neutrino masses as real as compared (and opposed) to any
other {\it possible} pattern, where the mixing angles can be anything one
wishes. {\it A priori}, no theoretical reason can determine which patterns is
real. In the quark sector, that the observed CKM mixing angles are almost
trivial and $\theta_{12}\gg \theta_{23}\gg\theta_{13}$ {\it completely} qualify
the observed hierarchical pattern of quark masses\cite{xuem}. In the lepton
sector, it seems unlikely to have exactly the same pattern for the following
observation: the hierarchical pattern of charged lepton masses is originated
dominantly from the hierarchical pattern of down quark masses (\ref{su5})
instead of a {\it possible} hierarchical pattern of the CKM-mixing angles, as
that in the quark sector. We have no theoretical reasons to preclude the 
possibilities of degenerate neutrino masses and large mixing angles. 

We expect that in eqs.(\ref{r1})-(\ref{r3'}), there exists a particular
solution giving the real pattern of neutrino masses and mixings in Nature, even
though we still have no power of making predictions. Nevertheless, these
relationships, originated from the purely theoretical stipulation, are
certainly facing all ongoing and future neutrino experiments.

\end{document}